\documentclass[12pt]{article}
\usepackage{amsmath}
\usepackage{graphicx}
\usepackage{amssymb}
\usepackage{geometry}
\usepackage{epsfig}
\begin{document}
\begin{center}
\begin{large}
\textbf{Reversal of magnetisation in Ising ferromagnet by the field having gradient}
\end{large}\end{center}
\vspace{1.5cm}
\begin{center}\emph{Abyaya $Dhar^{1}$ and Muktish $Acharyya^{2}$ }\\
\vspace{.4cm} 
\emph{Department of Physics, Presidency University,\\
86/1 College Street, Calcutta-700073, INDIA}\\
\vspace{.4cm}                                             
E-mail:$ ^1 $ abyaya93@gmail.com \\
E-mail:$ ^2 $ muktish.physics@presiuniv.ac.in\\
         
\end{center}
\vspace{1.5cm}
\textbf{Abstract:}We have studied the reversal of magnetisation in Ising ferromagnet by the field having gradient along a particular direction. We employed the Monte Carlo simulation with Metropolis single spin flip algorithm. The average lifetime of the metastable state was observed to increase with the magnitude of the gradient of applied field. In the high gradient regime, the system was observed to show two distinct region of up and down spins. The interface or the domain wall was observed to move as one increases the gradient. The displacement of the mean position of the interface was observed to increase with the gradient as hyperbolic tangent function. The roughness of the interface was
observed to decay exponentially as the gradient increases. The number of spin flip per site was observed to show a discontinuity in the vicinity of the domain wall. The amount of the discontinuity was found to diverge with the system size as a power law fashion with an exponent 5/3. 
\vspace{2.5cm}

\section{Introduction} The dynamical aspect of magnetization reversal in ferromagnetic system has been an active area of research. Particularly , for nucleation, the lifetime of metastable states \cite{guntonb} and decay modes are the main area of focus. Extensive simulation and experimental work has been done in last few decades. A work on nucleation in two three and four dimensions satisfying the classical predictions of Becker and Dorwing theory \cite{becker} with the help of heat bath dynamics was studied on Ising ferromagnetic system \cite{ma}. Kinetics of the nucleation phenomena, in the solid melt system and the dependences of lifetime of metastable states on magnetic field and system size, are well studied \cite{gunton, rikvold}. Magnetization switching of Heisenberg model for small ferromagnetic particle was also studied\cite{Unowak}. Macroscopic nucleation phenomena in continuum with long range interaction was observed \cite{macro}. FORC analysis of homogeneous nucleation in two dimensional kinetic Ising model was done \cite{forc}. Simulation of magnetization switching in nano particle systems  was gradually becoming popular\cite{hinzke}. Magnetization reversal modes in $L1_{0}$ Fe-Pt nano dots was studied with an atomistic modelling \cite{atom}. The effect of finite size on linear reversal mechanism was studied in a nano scale Fe-Pt \cite{ellis}. A decay of metastable phase for catalytic oxidation of CO was modelled and studied\cite{decay}. Reversal modes were simulated for Iron nano-pillar in an obliquely oriented field \cite{oblique}. Origin of asymmetric reversal modes in ferromagnetic/anti ferromagnetic multilayer system were also observed \cite{unowak2}. Previously there were many attempts to tune the nucleation time of the system. Ultra fast thermally induced magnetic switching in synthetic ferromagnet was studied \cite{ultra}. Heat assisted magnetization reversal was  also studied in ultra thin films by introducing a momentary, spatially localised input of energy in form of heat \cite{heat}. The reversal in Ising ferromagnet was studied by periodic pulse \cite{ma2} and very recently the nucleation was studied\cite{ma3} in Ising ferromagnet in the presence of a field having spatio temporal variation.

The response of a ferromagnet in the presence of a uniform magnetic field is well known.The lifetime of metastable state is extensively studied as a function of applied
magnetic field (uniform)\cite{ma}. The growth of nucleating clusters is also studied. 
This lifetime, plays an important role in the storage of magnetic devices. For 
practical purpose, the longevity of magnetic storage devices\cite{pira}, is
related to this metastable lifetime. To increase the longevity of storage devices, it is important to increase the metastable lifetime. {\it Can one increase
the lifetime of metastable state of a ferromagnet, by 
adjusting suitably the spatial variation of applied magnetic field ?}. 
To get the answer of this question, one may start from a simple feromagnetic 
model (e.g., Ising model) and minimal spatial variation of applied magnetic
field (having a gradient).
In this article, our main motivation, is
to study the statistics of the lifetime of metastable states 
and possible prolongation of this lifetime
by applying a field having a 
gradient (instead of uniform magnetic field).

In the present study, the above question is addressed. The manuscript is organized as follows: In the next section (section (2)) the model and simulation technique will be discussed. Next, the results from the numerical simulation are reported in section (3) and the paper ends with a concluding remarks mentioned in section (4).

\section{Model and Simulation Technique}
The Hamiltonian for Ising ferromagnetic system with nearest neighbour interaction and in presence of a spatially varying magnetic field can be represented as,
\begin{equation}
H=-J\sum_{<i,j>}{S_{i}}.{S_{j}}- \sum_{i}{h(i).S_{i}}
\end{equation}
Here $ S_{i}={\pm}1 $ are the Ising Spins. $ J (>0)$ is the ferromagnetic interaction strength and h(i) is the site dependent external magnetic field. Here , the form of this field is taken as
\begin{equation}
h(x)=g*x+c
\end{equation}
 where $ g=dh/dx$ is the gradient of the field. If initially $h_{l}$ and $h_{r}$ are the given field on the left boundary and the right boundary of the lattice respectively, g can have a form like $g=(h_{r}-h_{l})/L $ and c will be $h_{l}$. Here all the magnetic fields are measured in the units of J.\par
The range chosen here is $h_l=-0.5$ and $h_r=+0.5$. 
We have selected the range of values of fields at left and right sides in such a way
that the region of up and down spins in both sides are distinctly detectable. 

In the simulation, we start with a two dimensional lattice of size $ L{\times} L $  with {\bf open boundary conditions}. Initially the system is in perfectly ordered state where all the spins are up i,e $ S_{i} =1 $ ${\forall} $ i . In our simulation, we select each spin and calculate the energy required for spin flip ($S_{i}{\rightarrow}-S_{i}$) is $ {\Delta E} $. The flipping probability of the selected spin is determined by the  Metropolis Algorithm\cite{mcs}, 
\begin{center}
$P=Min(1,\exp{(-{\Delta}E/k_{B}T)})$
\end{center}
The temperature of the system T is measured in the unit $ J/k_{B} $, 
where $ k_{B} $ is the Boltzmann constant. Now a uniformly distributed 
(between 0 and 1) random number ($r$) is called. If this random number $r$ is less
than or equal to flipping probability $P$ then the spin was flipped. In this way,
$L^2$ such spin
were flipped in parallel updating scheme. This $L^2$ number of spin flips constitute a single
time step and defined as time unit (Monte Carlo Step per Spin or MCSS) in the problem
\cite{mcs}.  

We have chosen $L=300$ and kept it fixed throughout the study. The reason behind this choice is a compromise
between the affordable computational time and to have the clear observation of distribution of
metastable lifetimes. 

\vskip 1cm

\section{Results}
 In classical nucleation theory, depending on the value of applied 
uniform magnetic field, the multi droplet region and single droplet region are observed.
Figure-1a, shows a typical lattice morphology of a multi droplet region. The minimum time required to achieve negative magnetization from completely ordered state by applying a reversal field is called the nucleation time or the metastable lifetime of the ferromagnetic system. 
The variation of magnetization with ”time” is shown in Figure-2a.

Now, in the presence of a field which has a form like Equation (2), it is observed (in Figure-1b) that, instead of distribution of clusters, down spins grows from the boundary with higher value (absolute) of field. After nucleation a very rough interface is created which separates the regions positive and negative spins. Now an increase in amount of the gradient of magnetic field, the roughness of the interface( Figure-1c) decreases. Figure-1d shows that, for higher value of the gradient, the distinction is more prominent. For different gradients the variation of magnetization with time is shown in Figure 2. Figure-2b shows at lower gradient the variation is similar to case of that for steady magnetic field.

Then to check the spatial variation of density of down spins the length L (in the direction of gradient ) is divided into some strips of fixed width. At each strip the total 
number of down spins ( $S_{i}=-1$) is calculated. Dividing it by the total number of spins in that strip the density of down spins can be obtained. In a steady magnetic field in reverse direction the density of down spins should be constant, and just after nucleation it is expected to be something closer to 50 percent of the total spins (Figure-3a). But in case of a magnetic field which has a gradient, this is not constant. It start decreasing in a region and slowly becomes zero. This variation is shown in Figure-3b. As the gradient is increased gradually, the fall of the density of down spins becomes sharp (Figure-3c). And at higher gradient, it is similar to a step function which is clear indication of a 
distinct separation between the regions of negative and positive spins at the central line of lattice(Figure-3d).

 The spatial variation, of Number of spin flip per site, shows this interface is playing an important role in the dynamics of the system. For this measurement also we divided the length L (in the direction of gradient) into some strips of equal width and calculate the number of times the spin has flipped. Then dividing it by total number of spins in the strip we get the Number of spin flip per site. The results for steady magnetic field and that for the magnetic field with different gradients are shown in Figure-4. It may be noted that for higher values of the gradient, particularly at the neighbourhood of interface(in the central line of the lattice) the change in Number of spin flip is huge, so there might be a discontinuity(Figure-4d) as the system size ($L$) becomes infinitely large. 

Now a further increase in gradient, discontinuity increases rapidly (Figure-5a). To check whether the amount of the discontinuity is same for all the lattice sizes, the amount of the discontinuity is studied as a function of system size
($L$) (for a fixed gradient and temperature). This is shown in Figure-5b. Figure-5c shows that the discontinuity has a power law like variation with an exponent $\frac{5}{3}$. So in the limit of large gradient the central line separates two close neighbouring regions, namely hard (where the number of spin flip is low) and soft (where the number of spin flip is high).

To detect the exact position of any point on the the interface ,for each spin we check its 10 nearest neighbours on both sides. From Figure-2d we can say at the position of interface there will be down spins in the opposite direction of gradient (where the magnitude of the reversal field is high) and up spins in the direction of gradient (where the magnitude of the reversal field is low). When for a particular spin both the number of up (in the direction of gradient) and down(in opposite direction) neighbouring spins are equal to a certain tolerance level, we fix that spin position as the position of interface. We decided the tolerance after looking at the snapshot of the spin configuration. In a $300{\times}300 $ lattice we have given a tolerance of 70 percent for finding the position. Similarly, for whole lattice, the position of points of interface was obtained. The variation of average position with gradient of field also shows some consistent results. The average position grows upto a certain value and then is fixed to a value close to position of central line of the lattice Figure-6a.
This displacement of the mean position of domain wall\cite{domain} is 
observed to be a {\it hyperbolic tangent} function of the gradient (Figure-6a). The variation of roughness
 of the interface with gradient is shown in Figure-6b. It is observed that the roughness decreases exponentially (Figure-6b) with field gradient. The most
probale position of the interface was found to increase and becomes steady eventually, with the increase of the magnetitude of the gradient. This is shown in Figure-6(c).

The unnormalised distribution of the points of interface also consistent with the obtained results. If the gradient is increased the width of the distribution decreases(Figure-7) with an increase in most probable position. This increase of most probable( Figure-7c) and average value of the position of the interface 
indicates the motion of this interface upto a certain value of gradient.

 The site dependent magnetic field in lattice also increases the lifetime of ferromagnetic systems which is important for magnetic storage. The distribution of lifetime of such 50,000 identical systems in both steady and a spatially varying field are shown in Figure-8a. It shows that this gradient in magnetic field actually shifts the normal distribution and its most probable value to higher magnitude as compared to that observed the steady field case. The most probable(Figure-8c) and the average(Figure-8b) nucleation time was found to increases linearly with the magnitude of the gradients of applied field.

\section{Summary and Concluding Remarks}
The behaviour of the metastable state and its persistence, in an Ising 
ferromagnet are studied in the presence of a field (having a gradient), by extensive Monte Carlo simulation using Metropolis algorithm. The lattice morphology shows two distinct regions occupied by up and down spins, instead of showing distributed clusters as observed in the case of uniform (over the space) field. 
The most
probable value and the mean of the lifetime of the metastable states are observed to increase as the value of the gradient of field increases. This dependence is found to be linear. The interface or the domain wall was also observed to move as the gradient increases. Quantitatively, the displacement of the mean position of the interface increases as a hyperbolic tangent manner with the gradient of the field. The roughness of the interface decays exponentially as the gradient increases. The number of spin flip per site is also studied as a function of the gradient. This shows a discontinuity at the vicinity of the domain wall. The
amount of  discontinuity is found to diverge 
as the system size increases. This divergence is a power law with an exponent estimated equals to 5/3.

The distribution of lifetime of metastable states shows that the most probable
and the average lifetime increases as the gradient of the applied field 
increases.
The growth of metastable lifetime of Ising ferromagnet may be imagined as a possible increase of longevity of magnetic storage devices, if kept in a field having a gradient. It would be interesting to think of developing supporting theory as a generalisation of Becker-Doring droplet analysis.

\section{Acknowledgements}
We would like to thank Amitava Banerjee and Jayeeta Chattopadhay for their help in preparing the figures and for careful reading of the manuscript.

\newpage 
\begin{figure}[!tbp]
  \centering
  \begin{minipage}[b]{0.4\textwidth}
    \includegraphics[scale=0.6]{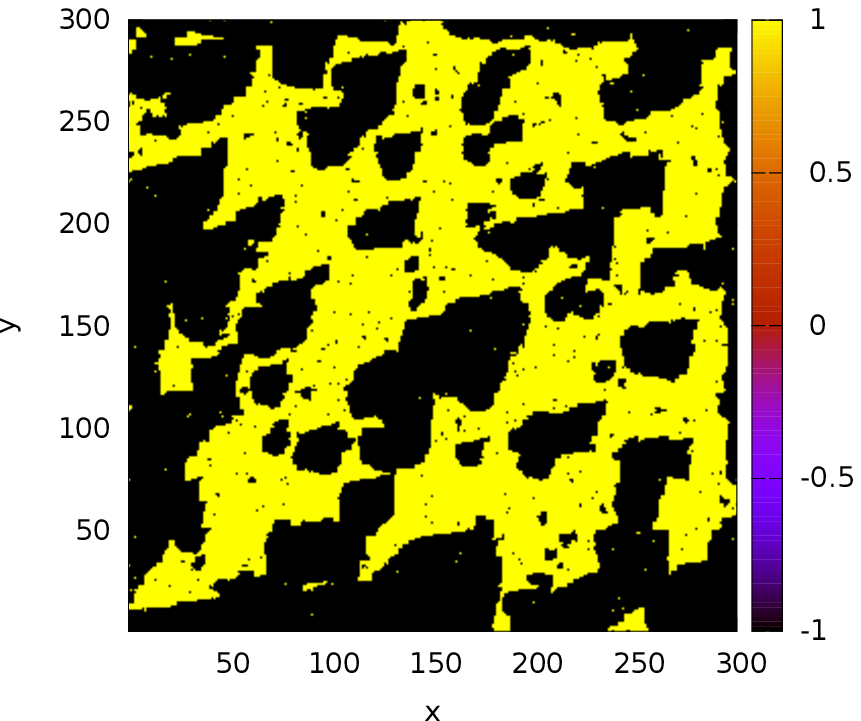}
  \end{minipage}
  \hfill
  \begin{minipage}[b]{0.4\textwidth}
    \includegraphics[scale=0.6]{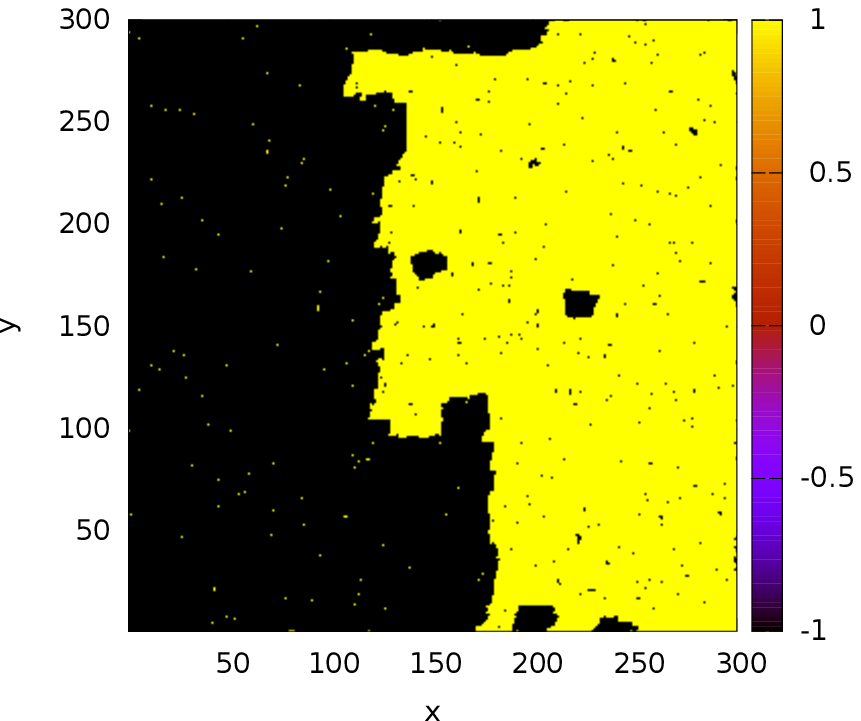}
\end{minipage}
\begin{minipage}[b]{0.4\textwidth}
    \includegraphics[scale=0.6]{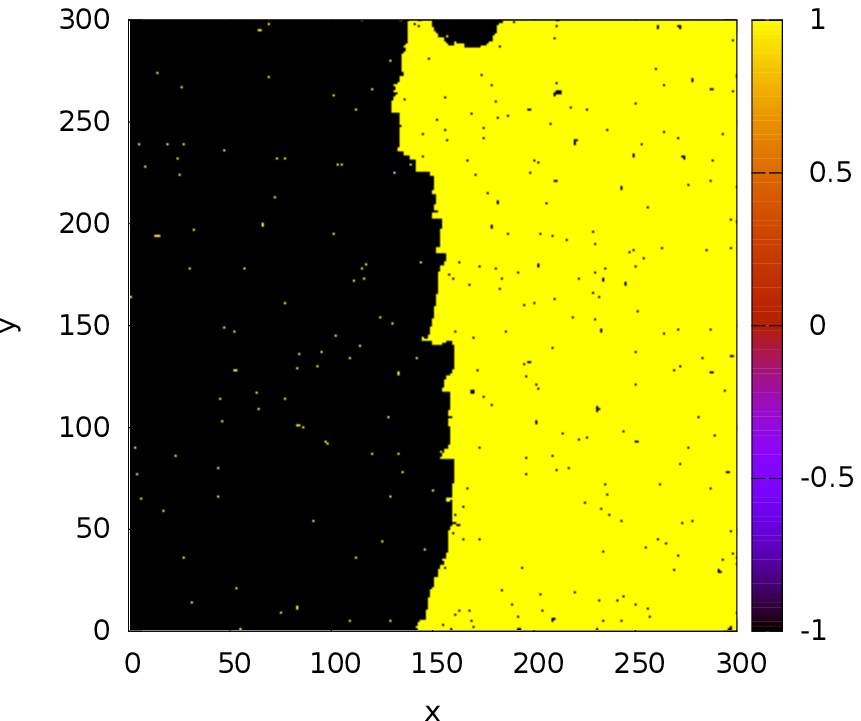}
 \end{minipage}
  \hfill
  \begin{minipage}[b]{0.4\textwidth}
    \includegraphics[scale=0.6]{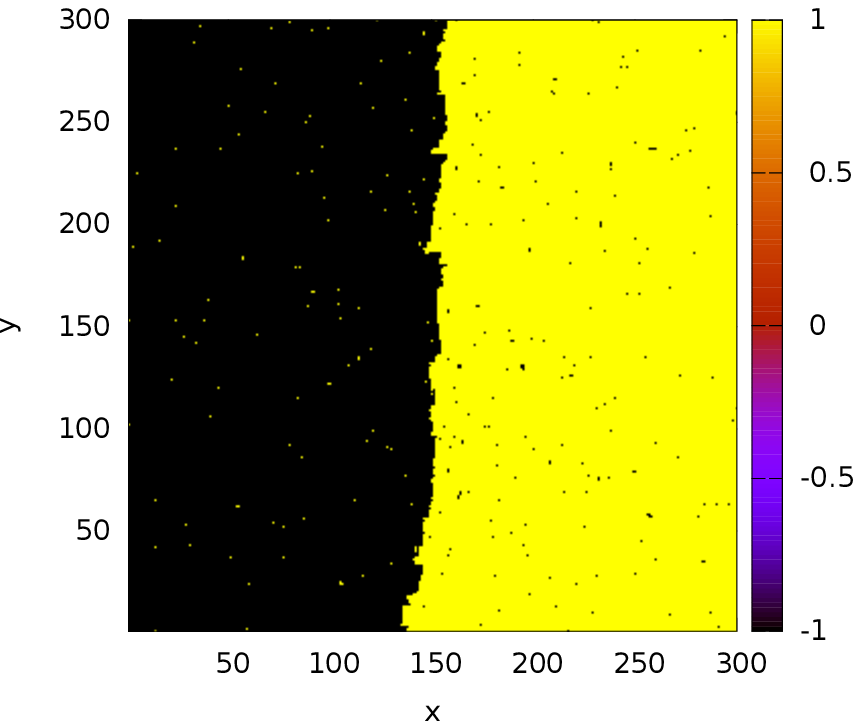}
\end{minipage}
\caption{Snapshots of spin configurations at nucleation time. {\bf Top left} a) The system at steady field h=-0.5.  {\bf Top right} b) The system with $h_{l}$=-0.5 and $h_{r}$=-0.2. {\bf Bottom left} c) The system with $h_{l}$=-0.5 and $h_{r}$=0.0. {\bf Bottom right} d) The system with $h_{l}$=-0.5 and $h_{r}$=0.2. Where $h_{l}$ and $h_{r}$ are the magnetic field on left boundary and right boundary of the lattice respectively, they are measured in the unit of $J$. 
Here, in all cases, the temperature $T=1.4 J/k_{B}$.} 
\end{figure}
\newpage
\begin{figure}[!tbp]
  \centering

    \includegraphics[scale=0.6]{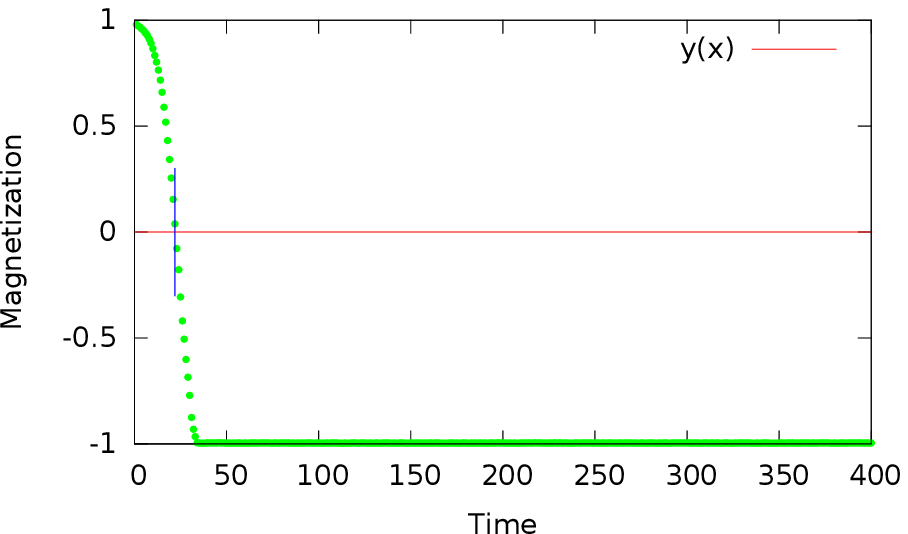}

    \includegraphics[scale=0.6]{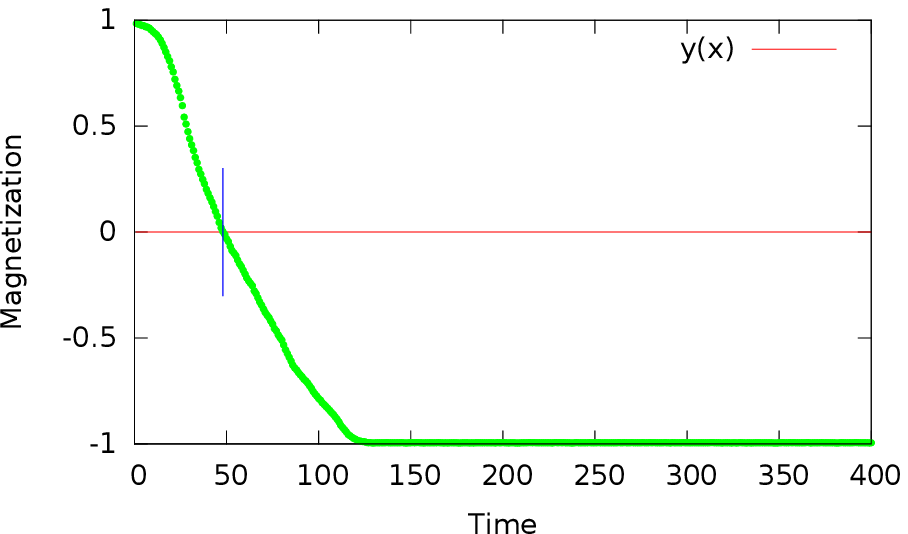}

    \includegraphics[scale=0.6]{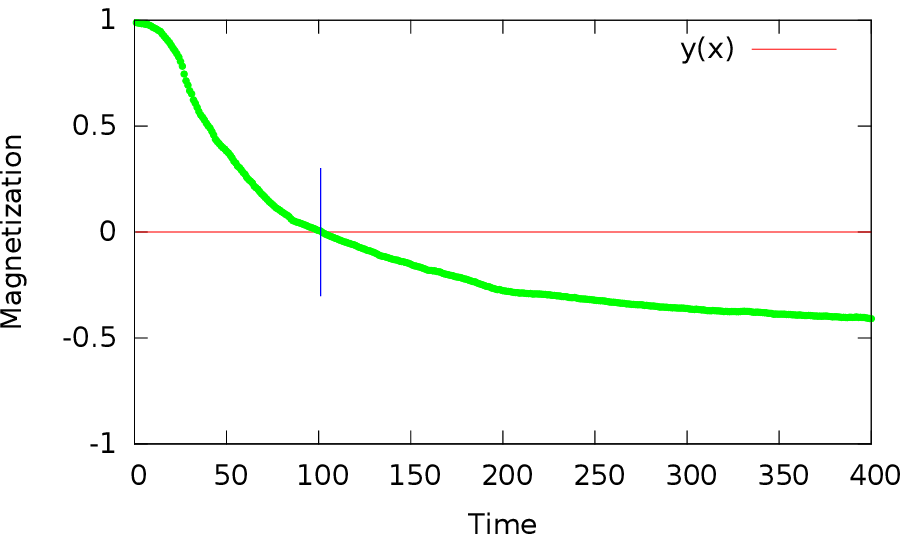}

    \includegraphics[scale=0.6]{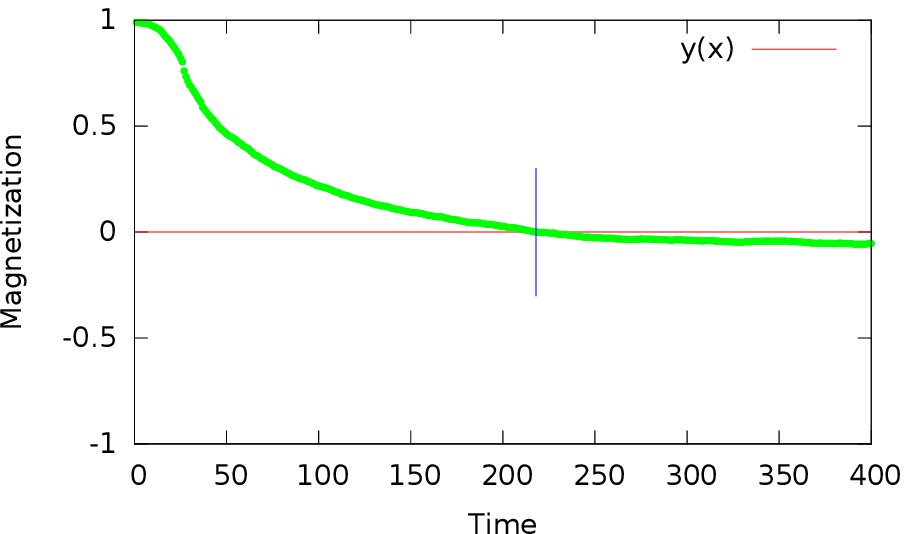}

\caption{Variations of Magnetization with time. {\bf Starting from top}  a) The system at steady field h=-0.5 b) The system with $h_{l}$=-0.5 and $h_{r}$=-0.2  c) The system with $h_{l}$=-0.5 and $h_{r}$=0.0  d) The system with $h_{l}$=-0.5 and $h_{r}$=0.2. Where $h_{l}$ and $h_{r}$ are the magnetic field on left boundary and right boundary of the lattice respectively, they are measured in the units of $J$. Here, in all cases, the temperature is $T=1.4 J/k_{B}$. The blue line marks the nucleation time.}
\end{figure}
\newpage
\begin{figure}[!tbp]
  \centering
  \begin{minipage}[b]{0.4\textwidth}
    \includegraphics[scale=0.6]{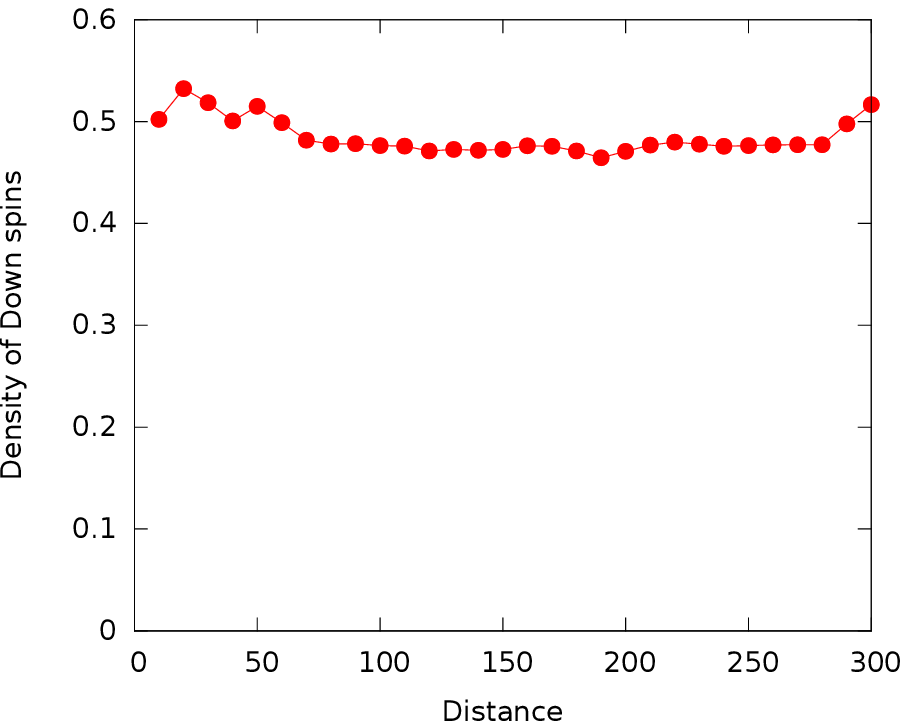}
  \end{minipage}
  \hfill
  \begin{minipage}[b]{0.4\textwidth}
    \includegraphics[scale=0.6]{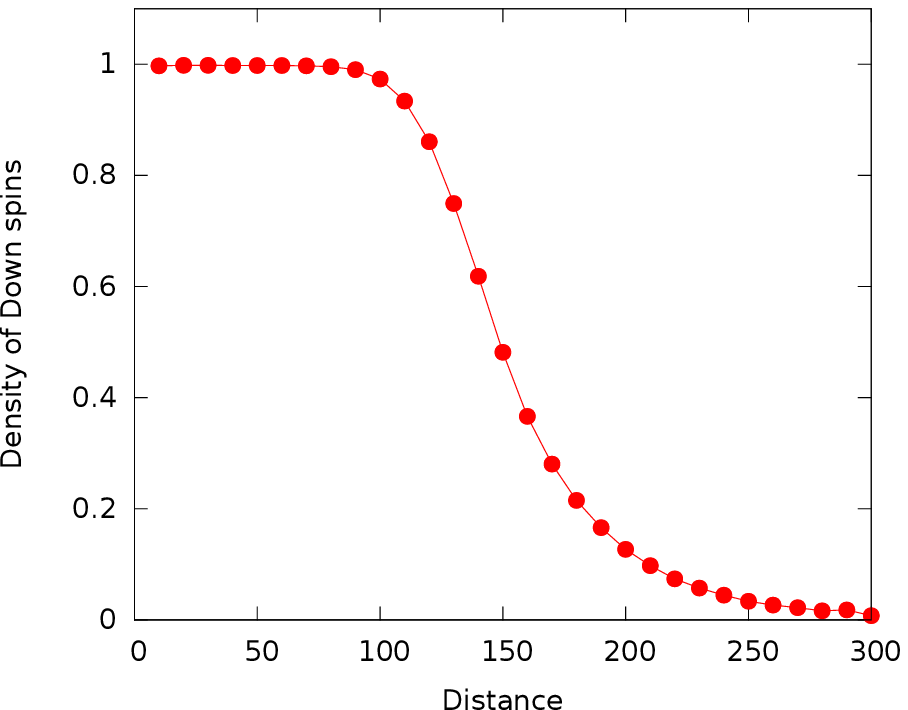}
\end{minipage}
\begin{minipage}[b]{0.4\textwidth}
    \includegraphics[scale=0.6]{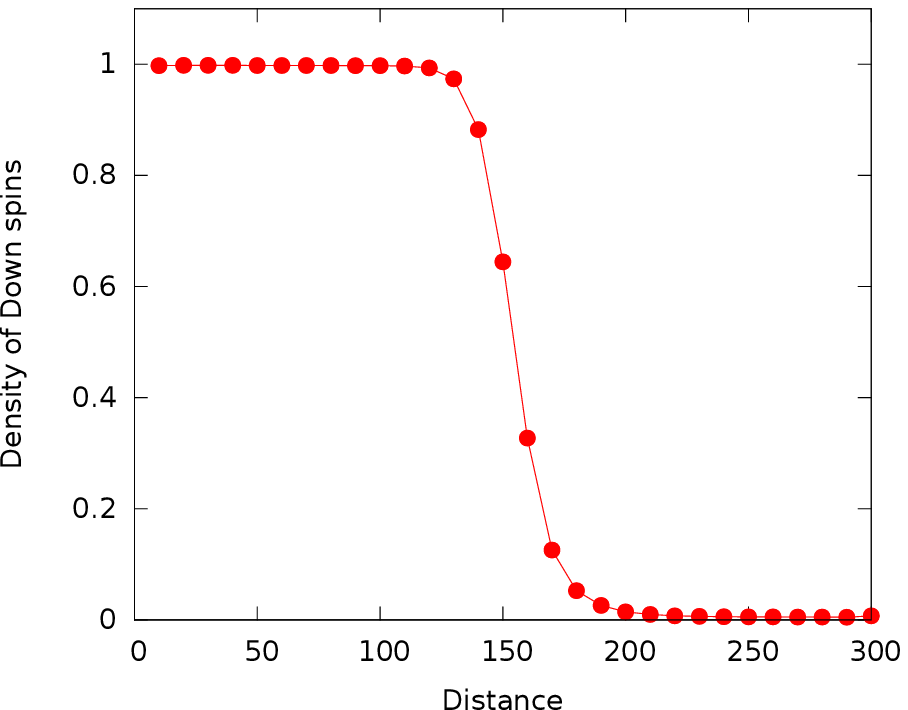}
 \end{minipage}
  \hfill
  \begin{minipage}[b]{0.4\textwidth}
    \includegraphics[scale=0.6]{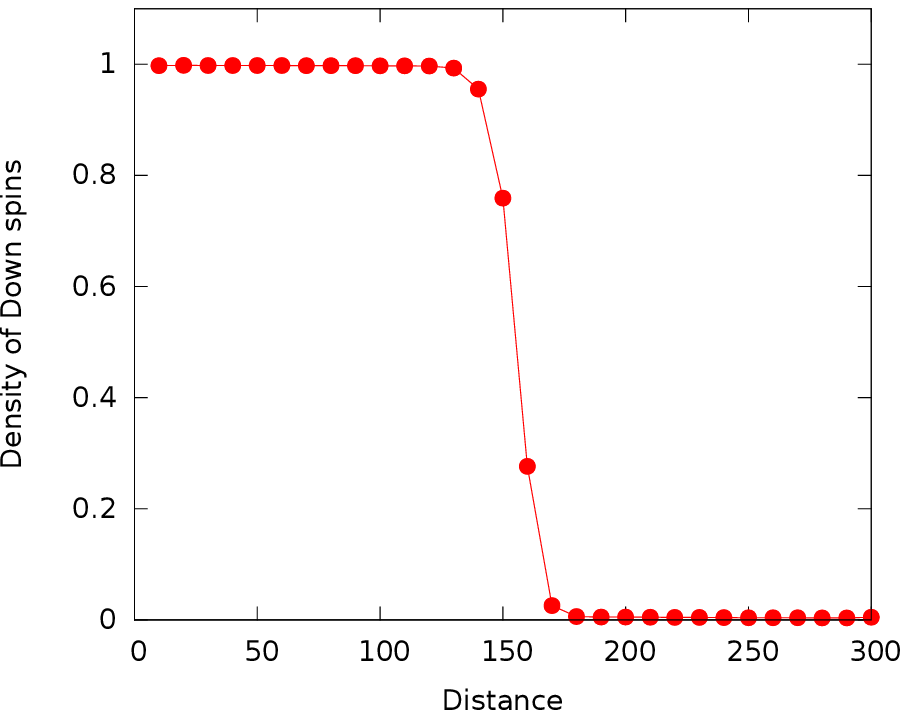}
\end{minipage}
 \caption{Spatial variations of density of down spins just after nucleation. {\bf Top left} a) The system at steady field h=-0.5. {\bf Top right} b) The system with $h_{l}$=-0.5 and $h_{r}$=-0.2. {\bf Bottom left} c) The system with $h_{l}$=-0.5 and $h_{r}$=0.0. {\bf Bottom right}  d) The system with $h_{l}$=-0.5 and $h_{r}$=0.2. Where $h_{l}$ and $h_{r}$ are the magnetic field on left boundary and right boundary of the lattice respectively, they are measured in the unit 
of $J$. Here, in all cases, the temperature is $T=1.4 J/k_{B}. $}
\end{figure}
\newpage
\begin{figure}[!tbp]
  \centering
  \begin{minipage}[b]{0.4\textwidth}
    \includegraphics[scale=0.6]{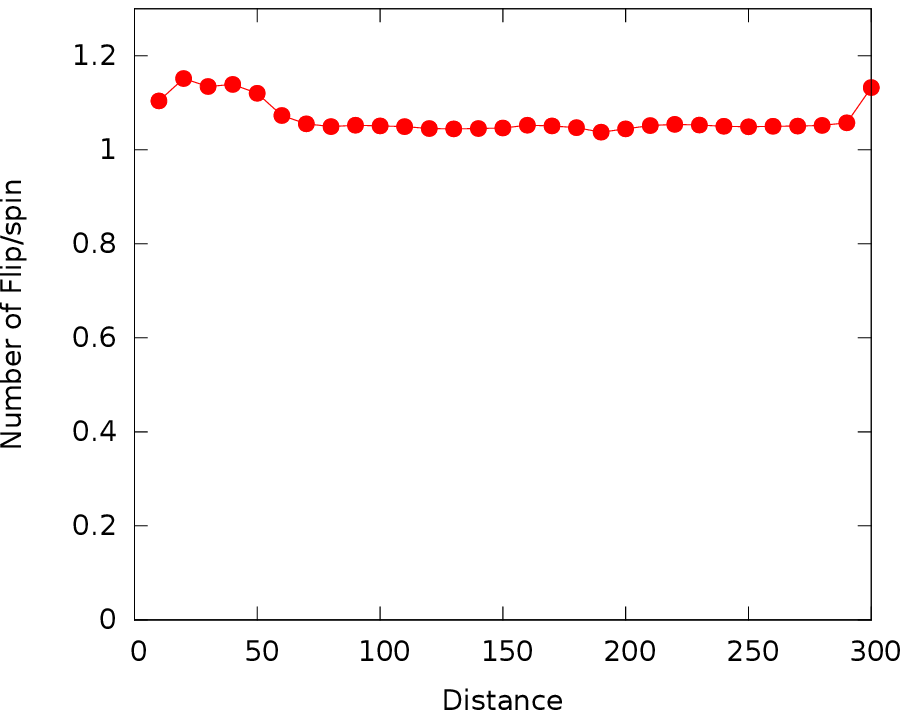}
  \end{minipage}
  \hfill
  \begin{minipage}[b]{0.4\textwidth}
    \includegraphics[scale=0.6]{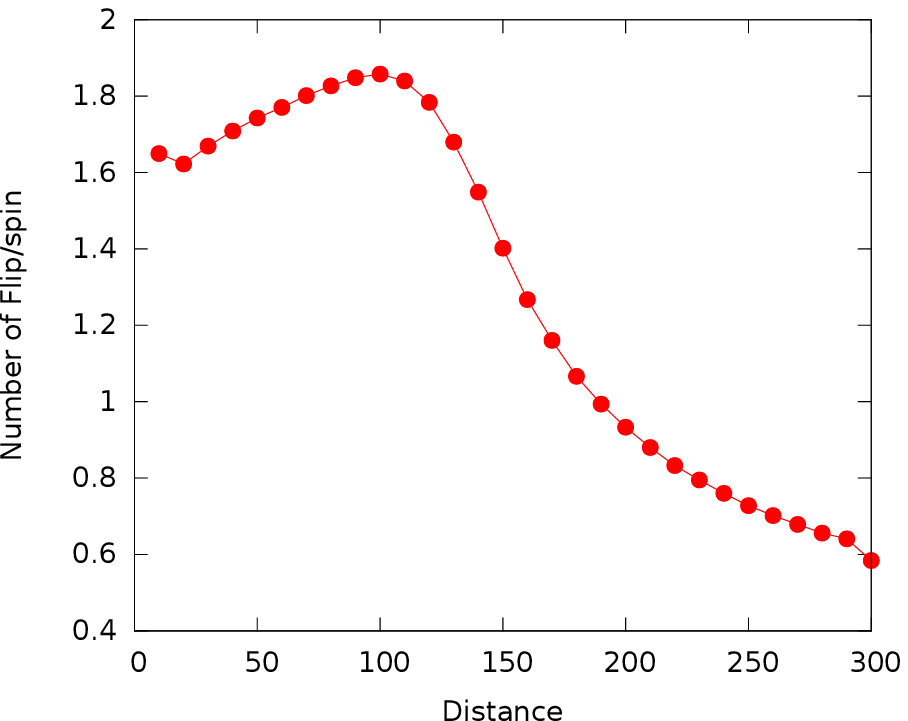}
\end{minipage}
\begin{minipage}[b]{0.4\textwidth}
    \includegraphics[scale=0.6]{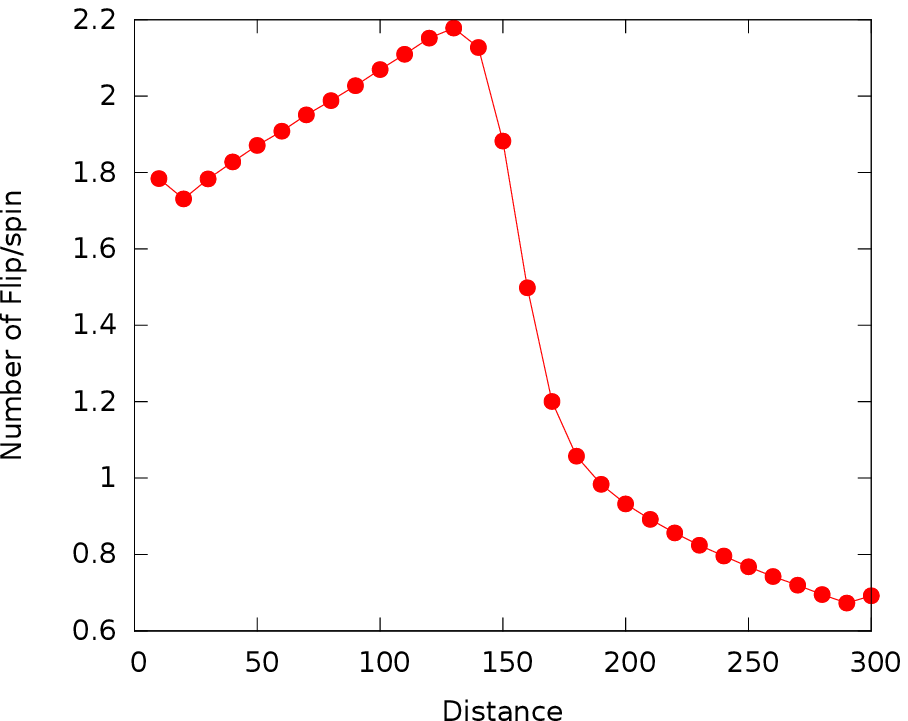}
 \end{minipage}
  \hfill
  \begin{minipage}[b]{0.4\textwidth}
    \includegraphics[scale=0.6]{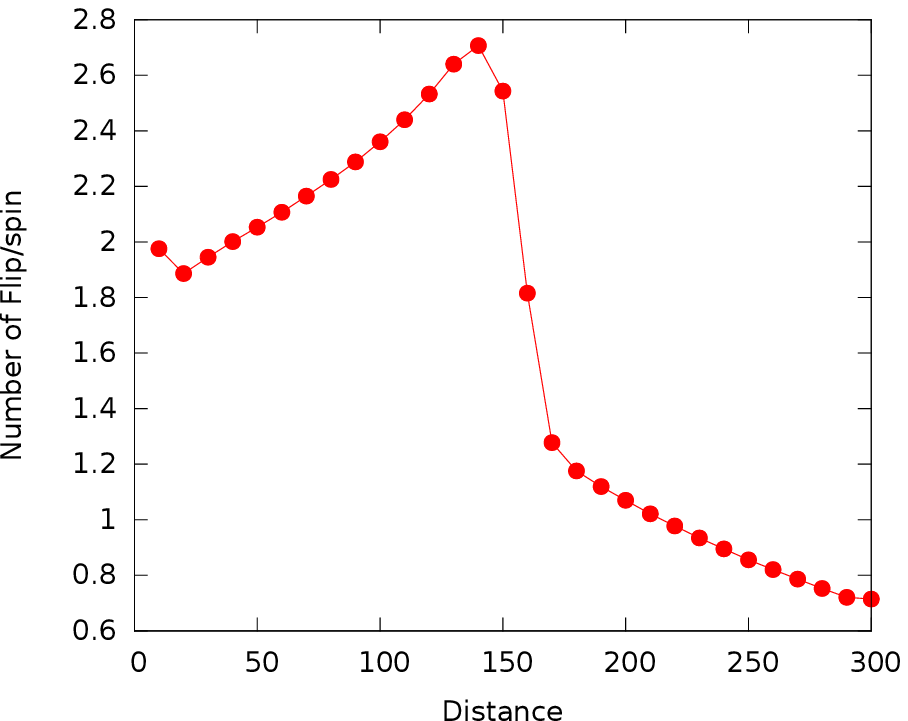}
\end{minipage}
\caption{Spatial variation of No of spin flip/spin just after nucleation.  {\bf Top left} a) The system at steady field h=-0.5. {\bf Top right} b) The system with $h_{l}$=-0.5 and $h_{r}$=-0.2. {\bf Bottom left}  c) The system with $h_{l}$=-0.5 and $h_{r}$=0.0. {\bf Bottom right}  d) The system with $h_{l}$=-0.5 and $h_{r}$=0.2. Where $h_{l}$ and $h_{r}$ are the magnetic field on left boundary and right boundary of the lattice respectively, they are measured in the unit 
of $J$. The temperature $T=1.4 J/k_{B} $, in all cases.}
\end{figure}
\newpage
\begin{figure}[!tbp]
  \centering
    \includegraphics[scale=0.6]{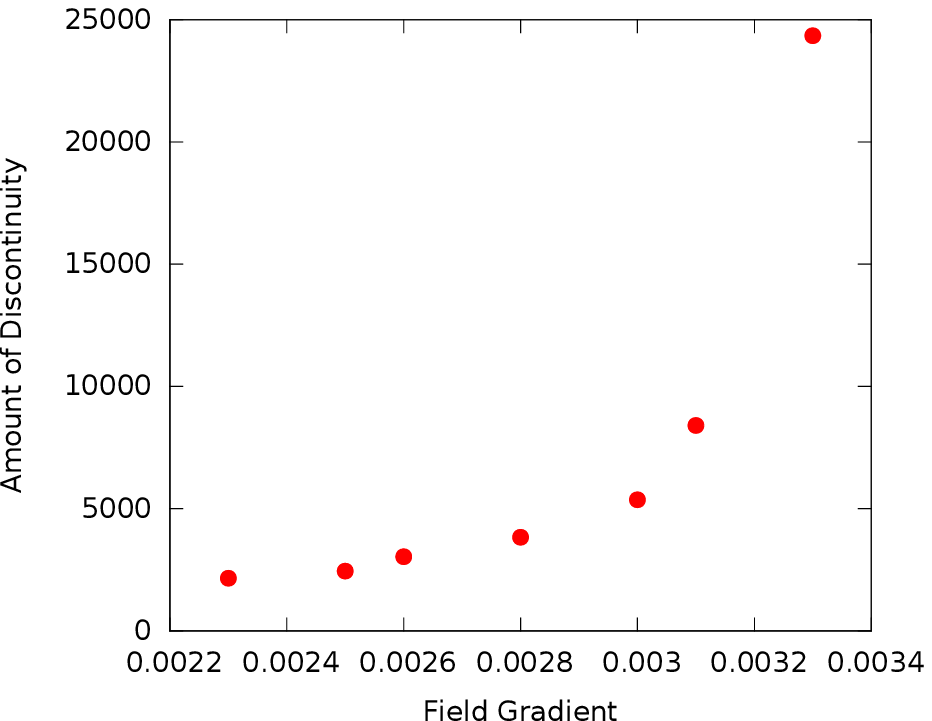}
    
\begin{minipage}[b]{0.4\textwidth}
    \includegraphics[scale=0.6]{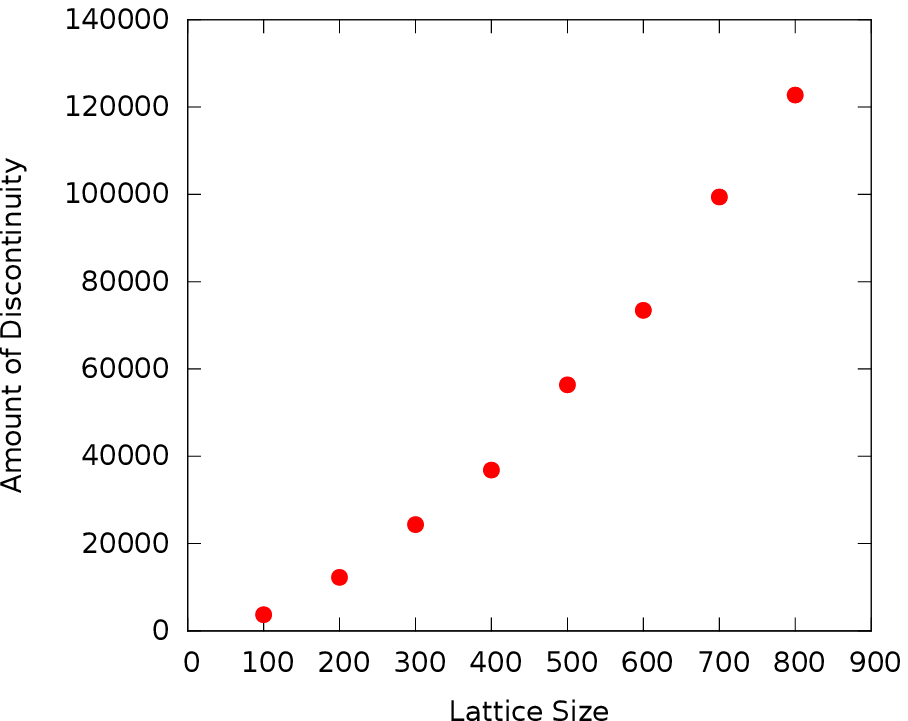}
\end{minipage}
\hfill
\begin{minipage}[b]{0.4\textwidth}
    \includegraphics[scale=0.6]{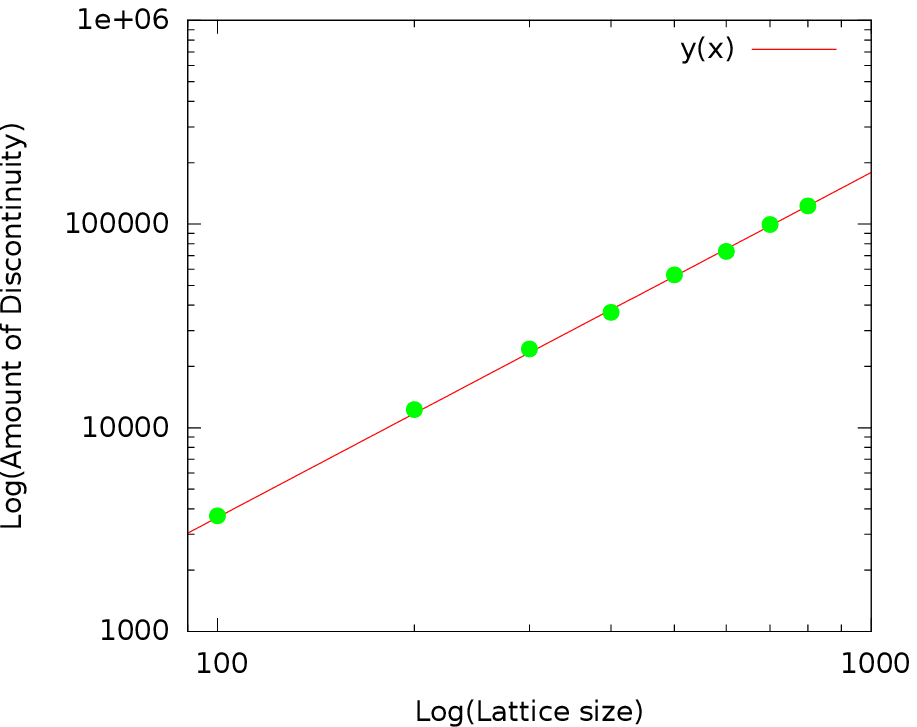}
\end{minipage}
\caption{ {\bf At the Top} a)Variation of the amount of discontinuity with applied gradient; Lattice size L=300 and  T= $ 1.4 J/k_{B} $. {\bf Bottom left} b) Variation of the amount of discontinuity with lattice size when $h_{l} $=-0.5 and $h_{r}$ =0.5. {\bf Bottom Right} c) Variation in the log scale of the amount of discontinuity with lattice size. $y(x)=1.4861x^{1.68}$,  when $h_{l} $=-0.5 and $h_{r}$ =0.5 }
\end{figure}
\newpage
\begin{figure}[!tbp]
  \centering
  \begin{minipage}[b]{0.4\textwidth}
    \includegraphics[scale=0.6]{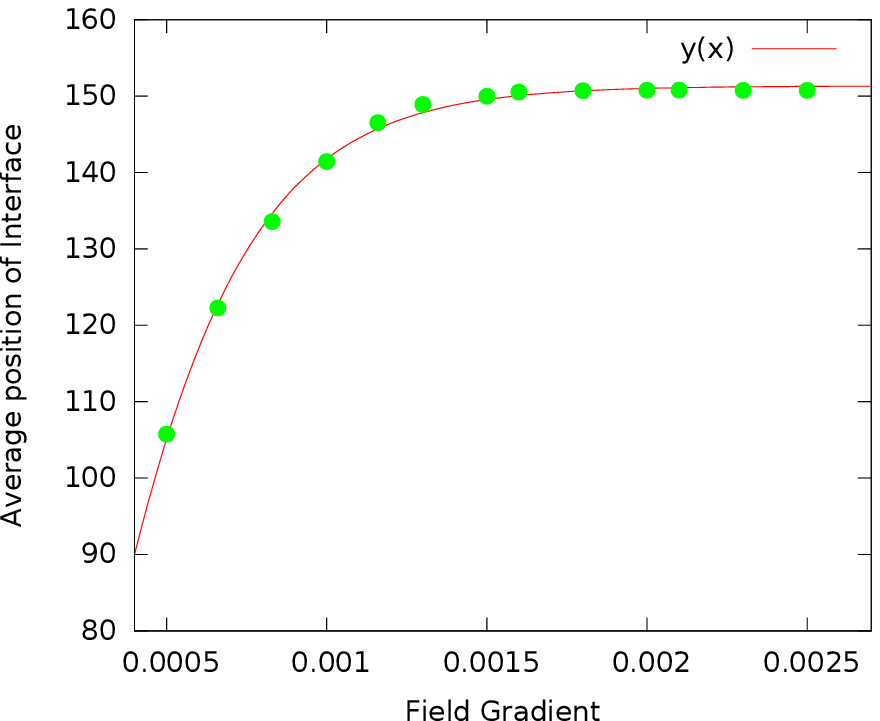}
  \end{minipage}
  \hfill
  \begin{minipage}[b]{0.4\textwidth}
    \includegraphics[scale=0.6]{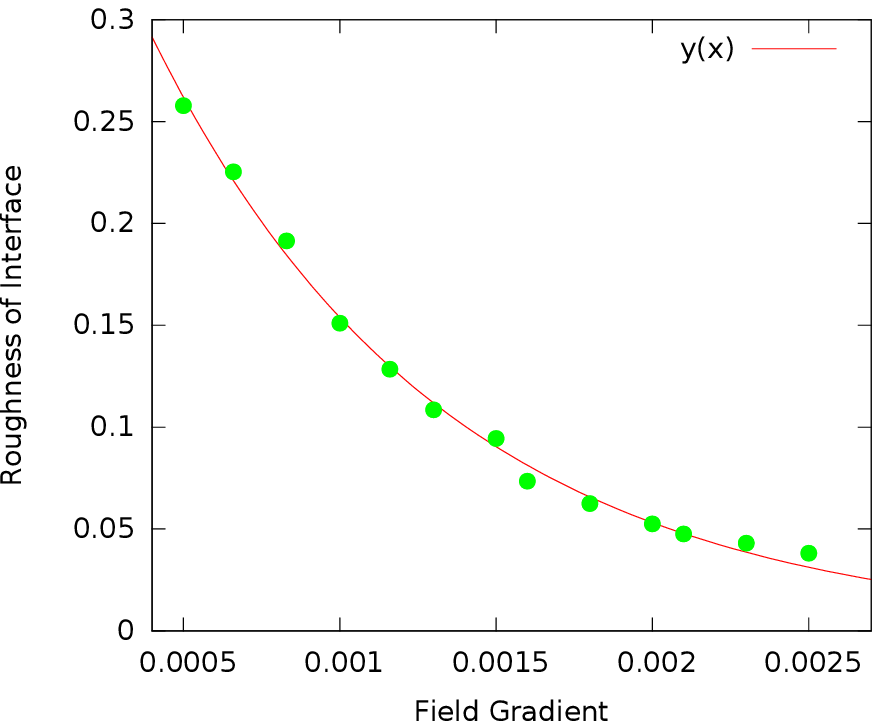}
\end{minipage}
\begin{minipage}[b]{0.4\textwidth}
    \includegraphics[scale=0.6]{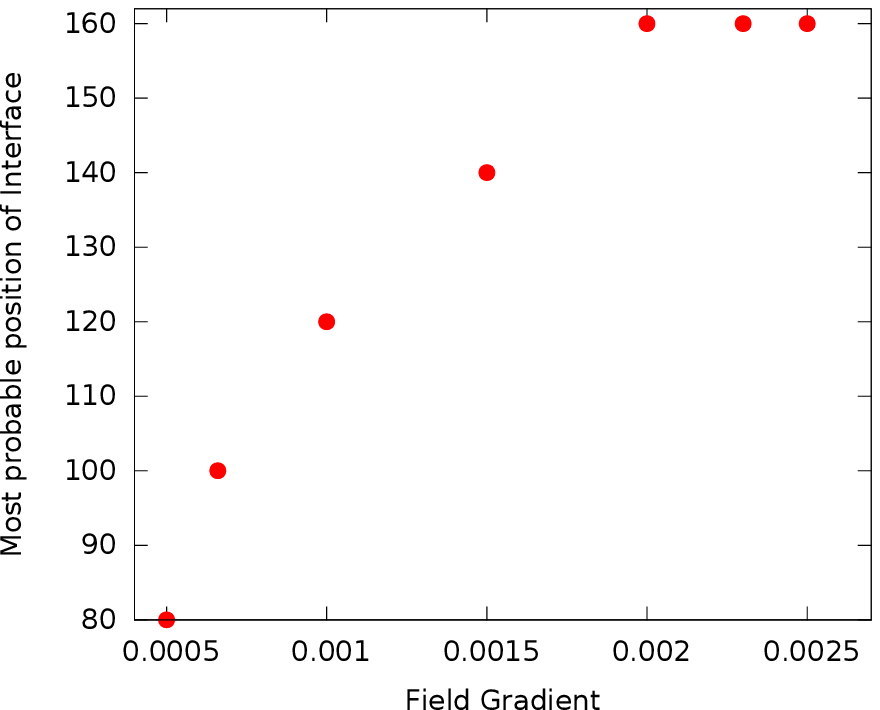}
 \end{minipage}
\caption{ {\bf Top left} a)Variation of Average position of the interface with gradient. The nature of variation is fitted with a curve $y(x)=151.329{\rm tanh}(1711.96x)$. {\bf Top right} b) Variation of roughness of interface with the gradient. The nature of the variation fitted with a curve $y(x)=0.445956{\rm exp}(-1063.83x)$. {\bf At bottom} c) The variation of most probable position of interface with field gradient. The temperature, $ T=1.4  J/k_{B} $, in all cases.}
\end{figure}
\newpage
\newgeometry{bottom=1.05cm}
\begin{figure}[!tbp]
  \centering
 
    \includegraphics[scale=0.6]{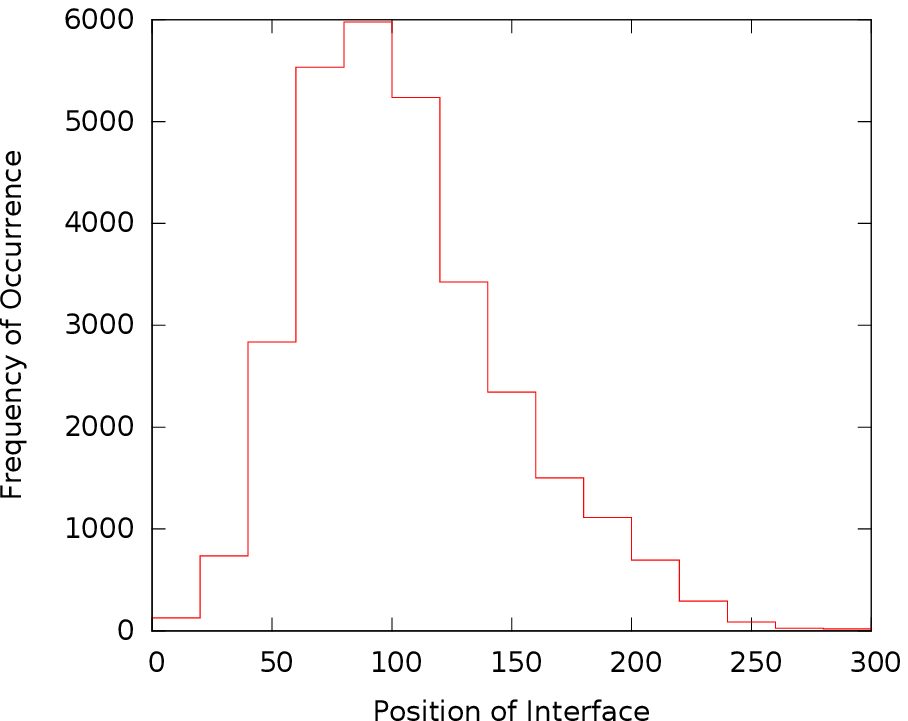}
  
    \includegraphics[scale=0.6]{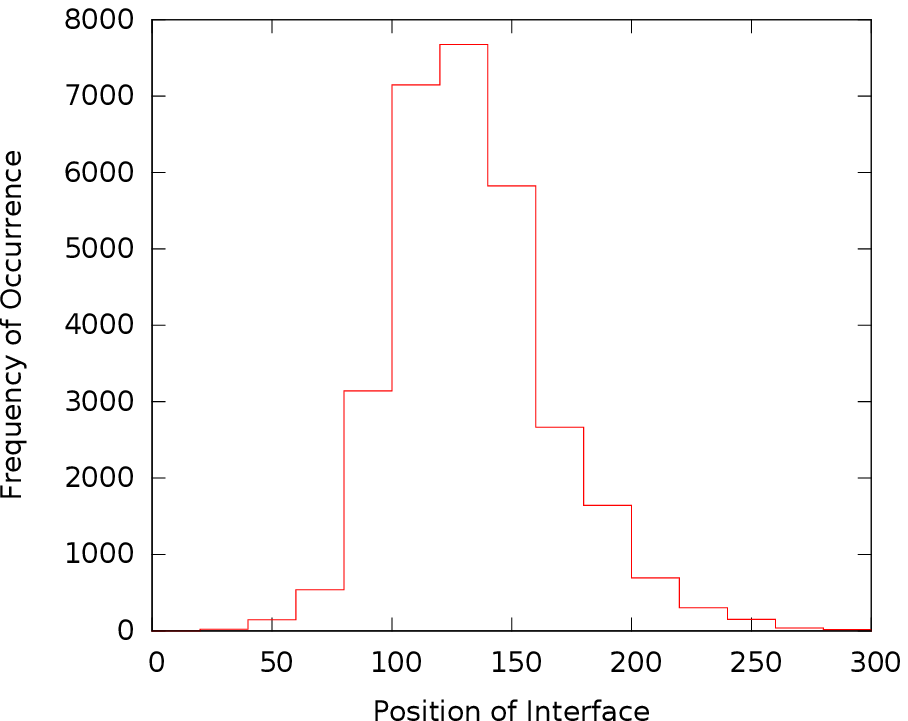}

    \includegraphics[scale=0.6]{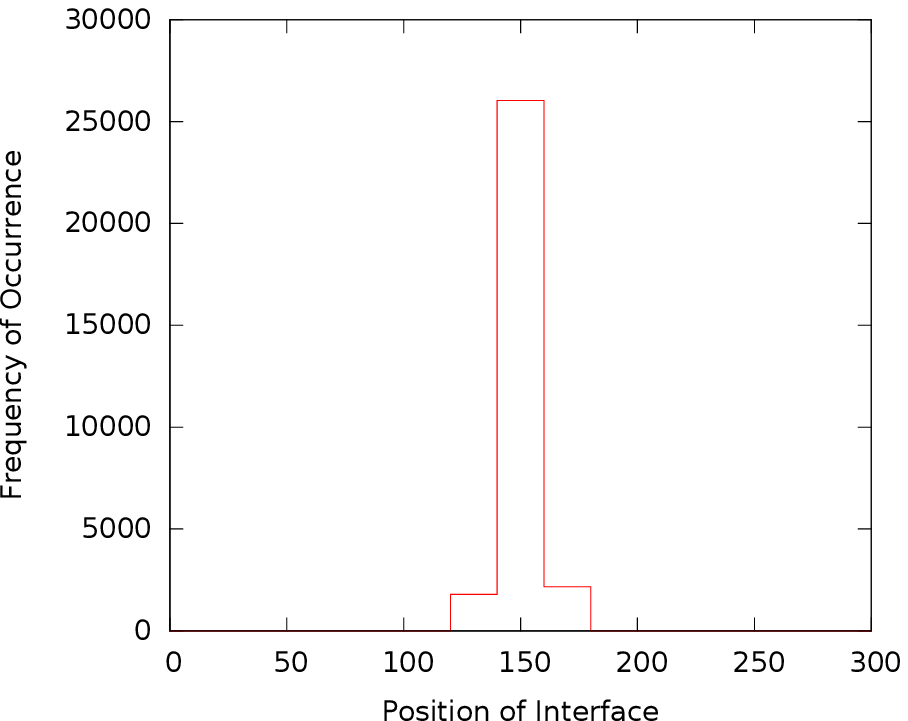}
\caption{ {\bf From the top}, Unnormalised distribution of position of interface a) $h_{l} $=-0.5 and $h_{r}$ =-0.3 b) $h_{l} $=-0.5 and $h_{r}$ =-0.15 c)  $h_{l} $=-0.5 and $h_{r}$ =0.15 , the temperature, $T=1.4  J/k_{B} $, in all cases. }
\end{figure}
\newpage
\begin{figure}[!tbp]
  \centering
    \includegraphics[scale=0.6]{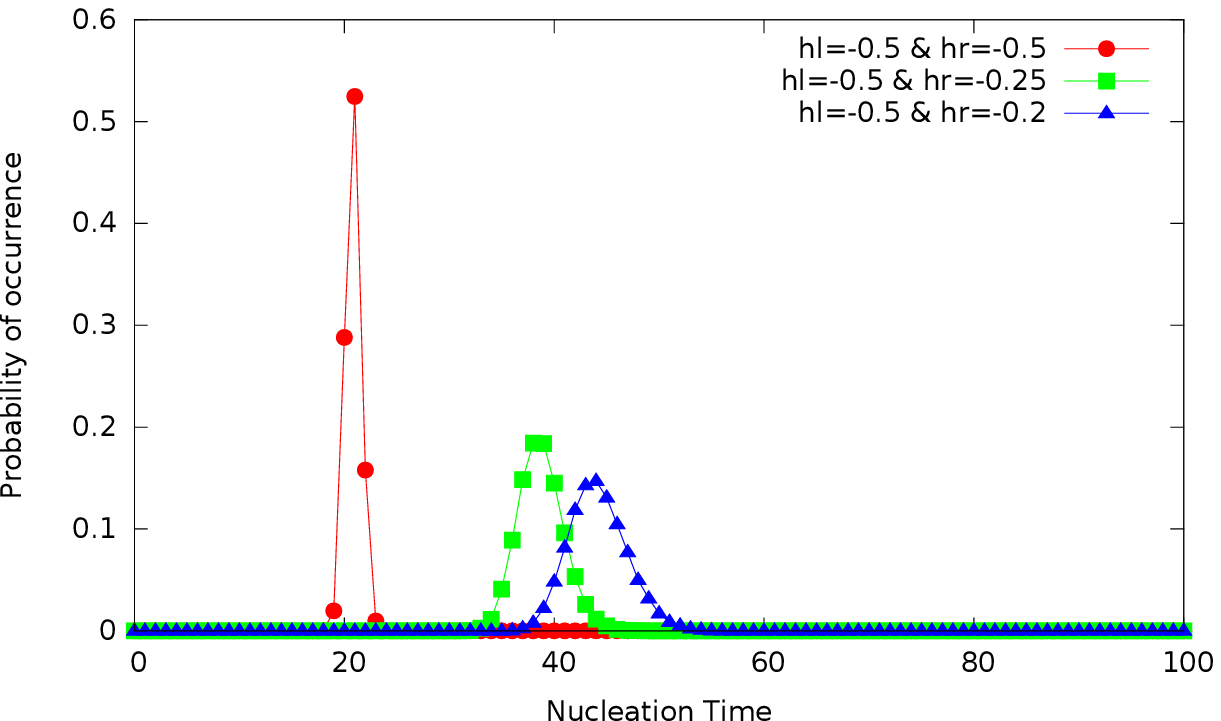}
 
    \includegraphics[scale=0.6]{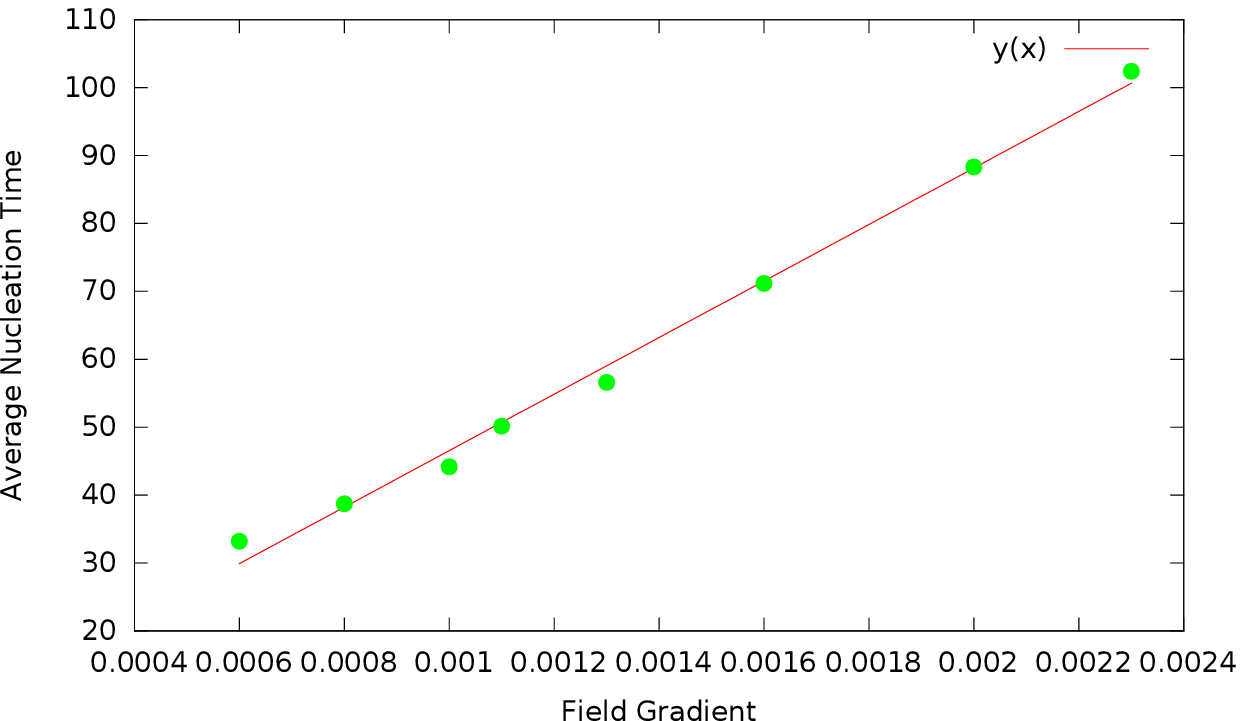}

    \includegraphics[scale=0.6]{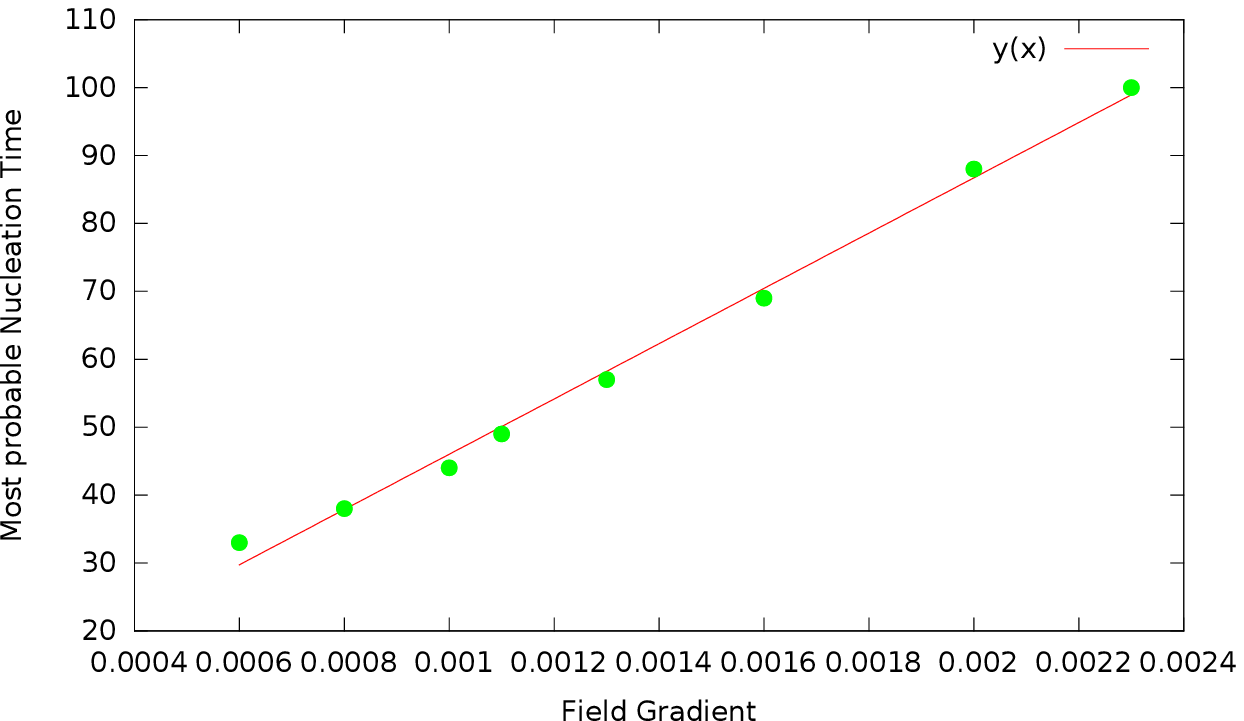}
 
\caption{ {\bf From the top}, a) The distribution of lifetime of 50,000 ferromagnetic systems for both steady magnetic field and field having a gradient, $h_{l} $ and $h_{r} $ denotes the field on left boundary and right boundary of the lattice respectively b) Variation of average nucleation time with field gradient. $y(x)= 41624.4x+4.91751$   c) Variation of most-probable nucleation time with field gradient $y(x)= 40707.3x + 5.30395$. The temperature $T=1.4 J/ k_{B} $, in all cases.}
\end{figure}

\newpage
\begin{center}{\bf References}\end{center}
\begin{enumerate}
\bibitem{guntonb} J.D Gunton, M Droz , \emph{Introduction to theory of Metastable and Unstable states (springer-verlag Berlin 1983)} 
\bibitem{becker} R. Becker and W. D¨oring, Ann. Phys. (Leipzig), 24 (1935) 719.
\bibitem{ma}M. Acharyya and D Stauffer,  European Physical Journal B, 5 (1998) 571
\bibitem{gunton}Martin Grant and J.D Gunton, Physical review B, 32 (1985) 11.
\bibitem{rikvold} P.A.Rikvold, H.Tomita, S.Miyashita and S.W.Sides, Phys. Rev. E, 49 (1994) 5080.
\bibitem{Unowak}D. Hinzke and U Nowak , Phys Rev B, 58 (1998) 265
\bibitem{macro}M Nishino , C. Euachesn , S. Miyasita , P.A Rikvold , K. Boukhelson and F Varmet, Scientific Reports, 1 (2011) 162 
\bibitem{forc}P.A Rikvold , M.A Novotny and D.T Robb, J Applied Phy, 97 (2005) 10E510
\bibitem{hinzke}D. Hinzke and U. Nowak, Phys. Stat. Sol, 189 (2002) 475
\bibitem{atom}J-W chantrell , U Atxitia , Richard F.L Evans , R.W Chantrell and C.H Lai, Phys Rev B, 90 (2014) 174415
\bibitem{ellis}M.O.A Ellis and R.W Chantrell, Applied Phys Letters, 106 (2015) 162407
\bibitem{decay}E. Machadao , G.M Buendia and P.A Rikvold , Phys Rev E, 71 (2005) 031603
\bibitem{oblique}S.H Thompson , G.Brown and P.A Rikvold , J Applied Phys, 97 (2005) 10E520
\bibitem{unowak2}B. Beckmann, U. Nowak and K.D Usadel, Phys Rev Letter, 91 (2003) 187 P01
\bibitem{ultra}R.F.L. Evans, T.A. Ostler, R.W. Chantrell, I. Radu and Th. Rasing , Applied Phys Letters, 104 (2014) 082410
\bibitem{heat}W. R. Deskins, G. Brown, S. H. Thompson and P. A. Rikvold, Phys Rev B, 84 (2011) 094431
\bibitem{ma2}M. Acharyya, Physica Scripta, 82 (2010) 065703
\bibitem{ma3}M. Acharyya, Physica A, 403 (2014) 94
\bibitem{pira} S. N. Piramanayagam and Tow C. Chong,  \emph{Development in Data Storage : Material Perspective, IEEE-Wiley Press (2011)}.
\bibitem{mcs} K.Binder and D.W Heermann , \emph{Monte Carlo Simulation in  Statistical physics, Second edition, Springer-Verlag (1992), Berlin}
\bibitem{domain}U. Nowak, IEEE Trans mag, 31 (1995) 4169
\end{enumerate}
\end{document}